\documentclass{PHYEAUTH}
\usepackage{graphicx}
\usepackage{amsmath}
\usepackage{amssymb}

\begin{document}

\begin{frontmatter}

\title{Critical conductance of the chiral 2d random flux model}

\author[address1]{Ludwig~Schweitzer\thanksref{thank1}}
and
\author[address2]{Peter Marko\v{s}}

\address[address1]{Physikalisch-Technische Bundesanstalt (PTB), Bundesallee 100,
  38116 Braunschweig, Germany} 

\address[address2]{Institute of Physics, Slovak Academy of Sciences,  84511
  Bratislava, Slovakia} 

\thanks[thank1]{Corresponding author. 
Ludwig.Schweitzer@ptb.de}

\begin{abstract}
The two-terminal conductance of a random flux model defined on a square lattice 
is investigated numerically at the band center using a transfer matrix
method.  Due to the chiral symmetry, there exists a critical point where the
ensemble averaged mean conductance is scale independent. We also study
the conductance distribution function which depends on the boundary
conditions and on the number of lattice sites being even or odd. 
We derive a critical exponent $\nu=0.42\pm 0.05$ for square samples of even width using 
one-parameter scaling of the conductance. This result could not be obtained previously 
from the divergence of the localization length in quasi-one-dimensional systems due to 
pronounced finite-size effects.
\end{abstract}

\end{frontmatter}

\section{Introduction}
The electronic transport properties of disordered materials depend essentially
on the physical symmetry of the system under examination. The well 
known standard symmetry classes---orthogonal, unitary, and symplectic---are 
appropriate for describing situations that occur in the presence or absence of 
time reversal symmetry, and broken spin rotational symmetry, respectively. 
Recently, an additional chiral symmetry has been attracted considerable interest 
for disordered two-dimensional systems \cite{BMSA98,AS99,MBF99,Cer01}. 
This special symmetry may appear if the underlying lattice is bi-partite
\cite{ITA94}, at which eigenfunctions can occupy only one of the two sub-lattices. 
In two-dimensional lattice models with diagonal disorder, the chiral symmetry 
is not observed so that for orthogonal and unitary symmetry all electronic
states are localized. However, for non-diagonal disorder situations, like the
motion of quantum particles subject to random magnetic 
fluxes as investigated in the present contribution, the eigenvalues occur in 
pairs $\pm \epsilon_i$, a characteristic feature of the chiral symmetry
\cite{ITA94}. Appertaining to the chiral symmetry is a
critical point in the center of the tight-binding band with a diverging
localization length at energy $E=0$ \cite{MW96,Fur99} and a scale
independent critical conductance.

\section{Model and Method}
We consider a two-dimensional tight-binding lattice model where the 
perpendicular random magnetic field is introduced via complex hopping terms.
These are chosen such that the magnetic flux per plaquette is given by the sum
of the random Peierls phases along the two bonds in the $z$-direction
$2\pi\phi_m=\alpha_{m,m+a_z}-\alpha_{m+a_x,m+a_z}$.
The corresponding Hamiltonian with nearest neighbor hopping is
\begin{eqnarray}\label{ham}
H &=&
-\sum_{m} \Big( c^\dag_{m+a_x}c^{}_{m}+c^\dag_{m-a_x}c^{}_{m} + \\
& &\ \textrm{e}^{i\alpha_{m,m+a_z}}c^\dag_{m+a_z}c^{}_{m}
+ \textrm{e}^{-i\alpha_{m,m-a_z}}c^\dag_{m-a_z}c^{}_{m}\Big)\nonumber
\end{eqnarray}
where the width $L$ ($x$-direction) and the length $L_z$ ($z$-direction)
of the sample is given in units of the lattice constant $a$.
The $c^\dag_{m}$ and $c^{}_{m}$ create or annihilate a Fermi
particle at the site $m$, respectively.  The random fluxes are distributed
uniformly, $-f/2 \le \phi_m \le f/2$, with probability density
$p(\phi_m)=1/f$, and the randomness is maximal for $f=1$ used in the
following. The average magnetic flux through the system is zero. 
 
We calculate the dimensionless two-terminal conductance \cite{ES81} for systems
of length $L_z$ via
\begin{equation}\label{gg}
g=\textrm{Tr}\{t^\dag t\}=\sum_i^{N} \frac{1}{\cosh^2(x_i/2)}.
\end{equation}
Here, $t$ is the transmission matrix with the $x_i$ parameterizing its 
eigenvalues. $N$ is the number of open channels.
Dirichlet boundary conditions are imposed in the transversal direction.

\begin{figure}[b]
\begin{center}
\leavevmode
\includegraphics[clip,width=0.75\linewidth]{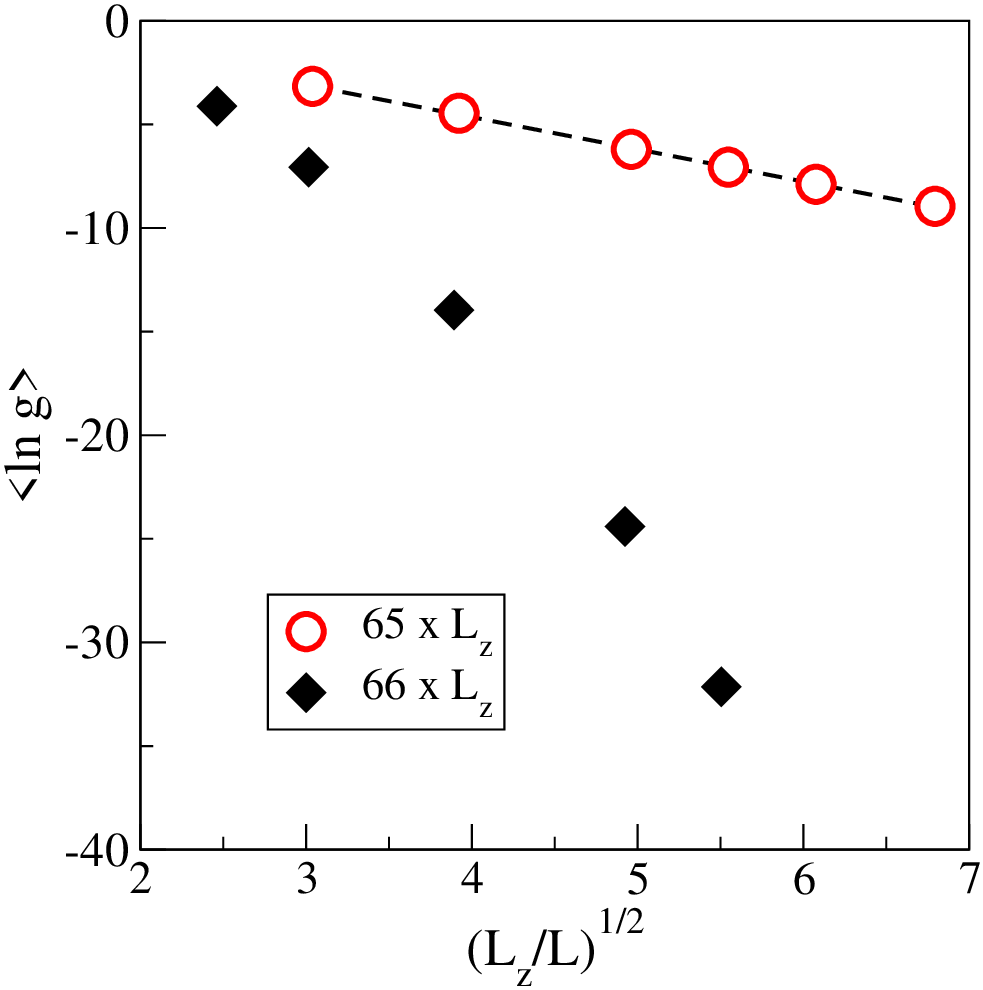}
\caption[]{The dependence of $\ln g$ on the square root of the aspect ratio
  $L_z/L$ (length to width).} 
\label{rmf-fig12}
\end{center}
\end{figure}

\begin{figure}[t]
\begin{center}
\leavevmode
\includegraphics[clip,width=0.97\linewidth]{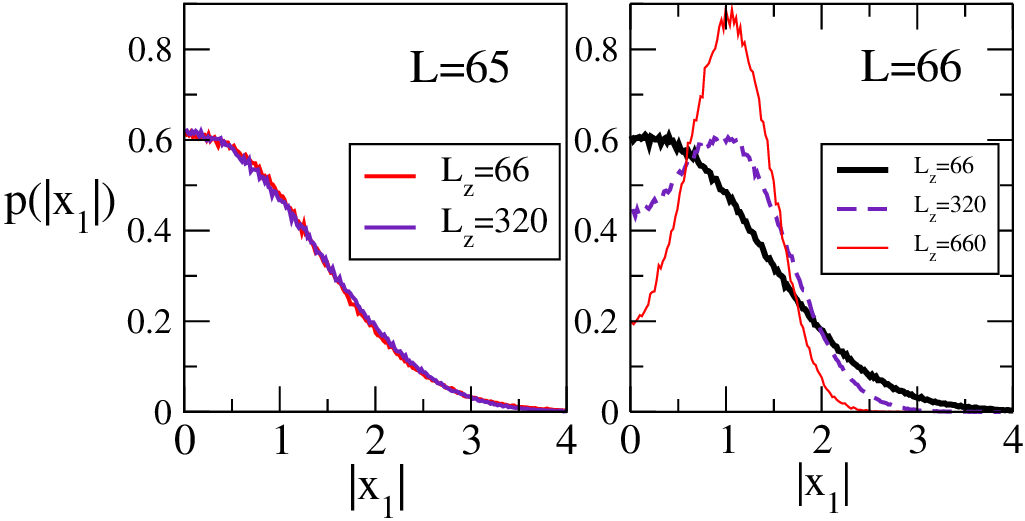}
\caption[]{The probability distribution $p(|x_1|)$ for odd ($L=65$) and even
  ($L=66$) sample widths.} 
\label{65-350a}
\end{center}
\end{figure}

\begin{figure}[b]
\begin{center}
\leavevmode
\includegraphics[clip,width=0.75\linewidth]{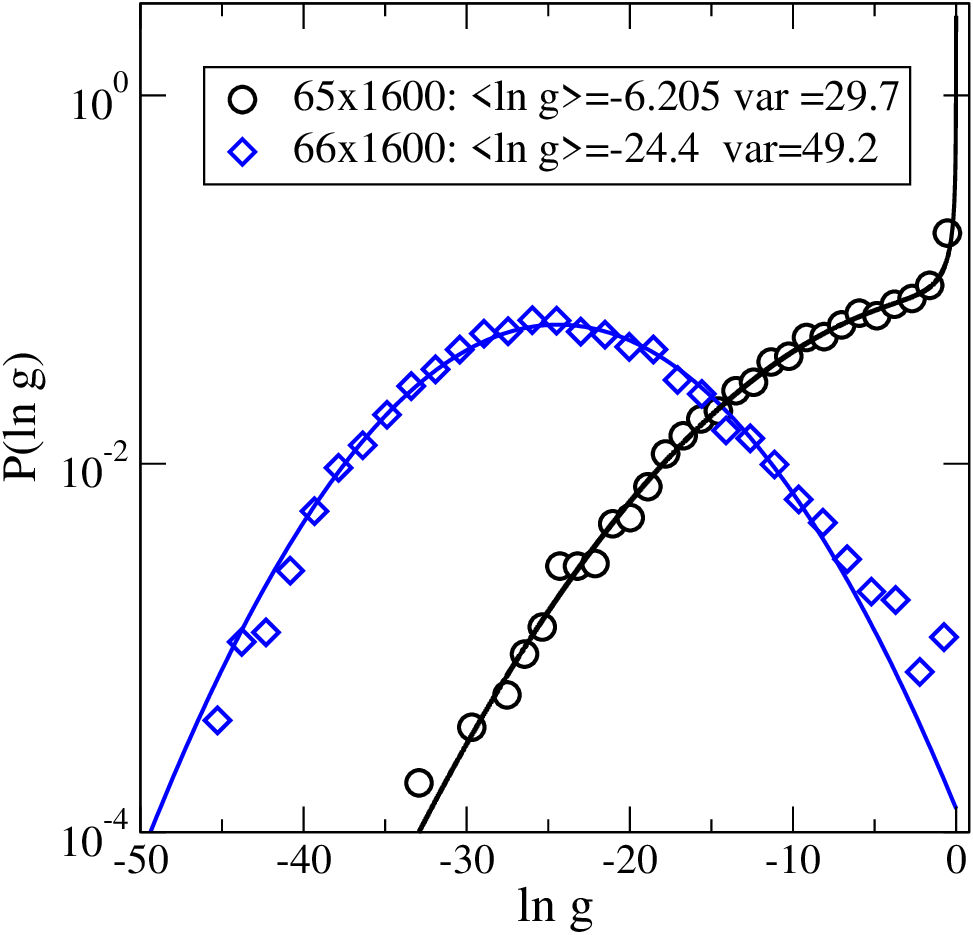}
\caption[]{The distribution of $\ln g$. The solid lines are given by the
  function (\ref{fit}). 
  Note the divergence of $p(\ln g)$ for $\ln g\to 0$ when $L$ is odd.} 
\label{rmf-plng-1}
\end{center}
\end{figure}

\section{Conductance}
Fig.~\ref{rmf-fig12} confirms the theoretical prediction \cite{MBF99}
that for quasi one-dimensional systems (Q1D), $\langle\ln g\rangle$ decreases
much slower for $L$ being odd 
than for the same system with $L$ even. This is easy to understand from
Eq.~(\ref{gg}) and from the mean values of the parameters $x_i$. In systems
with chiral symmetry, the $x_i$ assume both positive and negative
values. When ordered in absolute values, $|x_1|\le |x_2|\le \dots$, it follows
from symmetry reasons that $\langle x_1\rangle=0$ for $L$ odd, but $\langle
x_1\rangle = - \langle x_2\rangle\ne 0$  when $L$ is even. 
This is confirmed  in Fig. \ref{65-350a} which shows the probability distribution
$p(|x_1|)$ both for $L$ odd and even.  For $L$ even, $p(|x_1|)$ converges to
the Gaussian distribution with non-zero mean when the length
$L_z$ of the system increases, but $p(|x_1|)$ remains half-Gaussian for $L$ odd. 

The different form of the distribution of $x_1$ influences the transmission 
through very long Q1D systems and thus the conductance. Fig.~\ref{rmf-plng-1} 
shows the distribution of the logarithm of the conductance $p(\ln g)$.   
In very long Q1D systems, the conductance is determined by the first channel,
$g = \cosh^{-2}(x_1)$, so that
\begin{equation}\label{plng}
p(\ln g)=\int dx_1 p(x_1)~\delta(\ln g - 2\ln\cosh x_1).
\end{equation}
We find
\begin{equation}\label{fit}
p(y=\ln g)=
\frac{2}{\sqrt{2\pi\sigma}}\coth(\tilde{x}/2)
  \exp(-\tilde{x}^2/2\sigma), 
\end{equation}
where $\tilde{x}=2\ln[e^{-y/2}+\sqrt{e^{-y}-1}]$, and $\sigma=\langle
x_1^2\rangle$.
For $L$ even, $p(x_1)$ is Gaussian with a non-zero mean value.
Consequently, $p(\ln g)$ is Gaussian, too. The singularity of
$\coth(\tilde{x})$ in Eq.~(\ref{fit}) is eliminated by the exponential
decrease of $p(x_1)$. However, for $L$ odd,
$p(x_1)$ possesses a maximum at $x_1=0$ so that $p(\ln g)$ is
singular for $\ln g\to 0$. Since this singularity survives for any length of
the system, there is  a non-zero probability to obtain a sample with
conductance $g$ of order of unity. This explains the absence of exponential
localization in these systems, we find $\langle g\rangle\sim L_z^{-1/2}$ for
$L$ odd instead
as shown in Fig.~\ref{rmf-fig12}.

\section{Scaling of the conductance}

\begin{figure}[t]
\begin{center}
\leavevmode
\includegraphics[clip,width=0.9\linewidth]{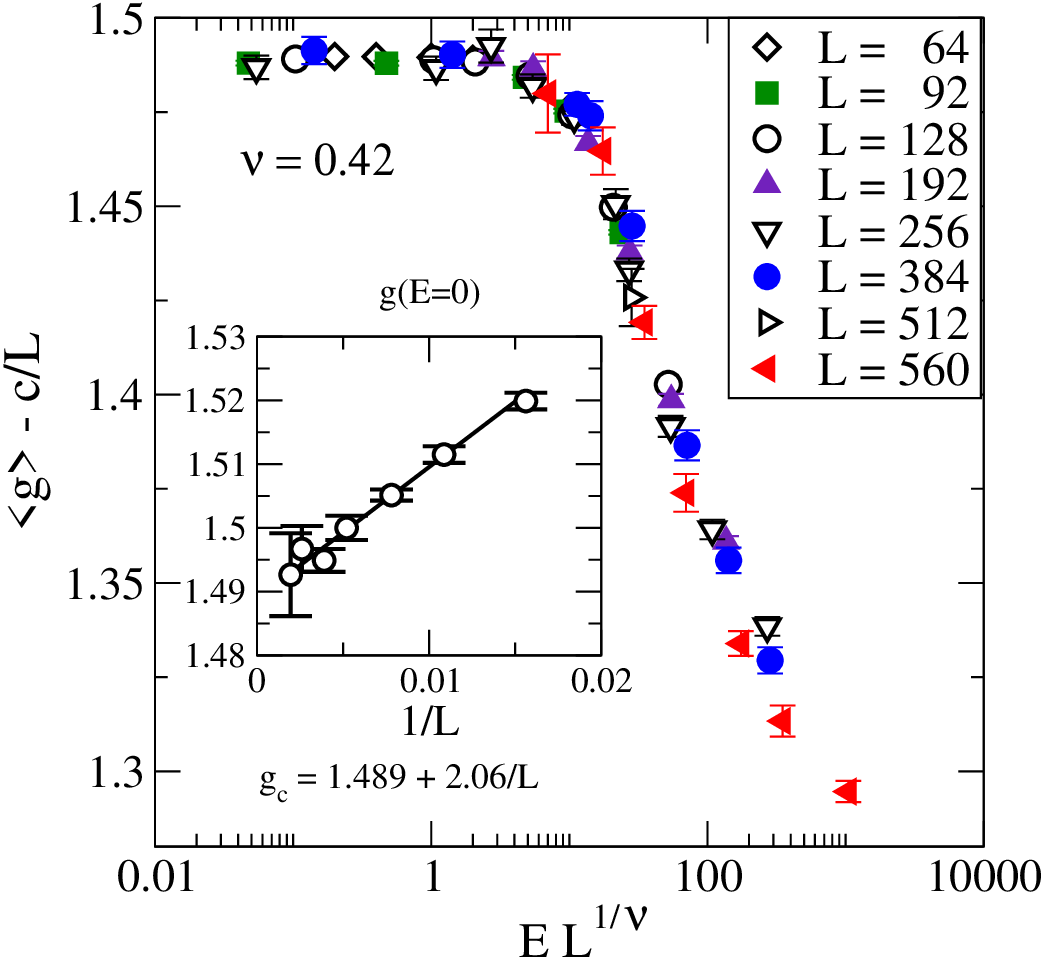}
\caption[]{The scaling of the conductance in the vicinity of the band center.
The inset shows the finite size corrections at the critical point. After subtracting
the irrelevant scaling term $2.06/L$ from the conductance, all data with 
$|E|< 0.0005$ scale to a universal curve.} 
\label{g-scal}
\end{center}
\end{figure}

Single parameter scaling theory requires that the conductance is a function of only
one parameter in the vicinity of the critical point,
\begin{equation} \langle g(E,L)\rangle = F\left[L/\xi(E)\right].\end{equation}
Here $\xi(E)$ is a correlation length which diverges as $\xi(E)\propto |E|^{-\nu}$.
In Ref.~\cite{MS07} we have analyzed the scaling of the smallest 
Lyapunov exponent of Q1D systems with \textit{odd} width and found the
critical exponent $\nu\approx 0.35\pm 0.03$. As also shown in 
Ref.~\cite{MS07}, no scaling analysis of the Lyapunov exponent is possible for
$L$ even because of strange finite size effects. Therefore, we verify in this
paper the one parameter scaling for 2D square samples $L\times L$ with even $L$ 
using the mean conductance (Fig. \ref{g-scal}). 
First, we subtract from the conductance the finite size correction, given by
the irrelevant scaling term $c/L$,
\begin{equation} \langle g(E=0,L)\rangle=g_c-c/L,\end{equation}
where $c=2.06$ and $g_c = 1.489 \pm 0.001$.
Then, the data scale to the universal curve, $\langle g\rangle
-c/L=F\left[EL^{1/\nu}\right]$ with 
a critical exponent $\nu=0.42\pm 0.05$, consistent with our recent result for
the divergence of the localization length in samples having odd width \cite{MS07}.

\section{Conclusion}
The two-terminal conductance of a random flux model has been investigated
numerically. It shows  critical behavior at
the band center due to the chiral unitary symmetry of the system. The critical
exponent obtained from a finite size scaling analysis agrees with the one
that governs the divergence of the localization length. These results
support the notion of universality at the chiral critical point.

PM thanks Grant from APVV, project n. 51-003505, and VEGA, project n. 2/6069/26.


\end{document}